\documentclass[preprint,aps,showpacs]{revtex4}

\usepackage{amsmath}
\usepackage{graphicx}
\usepackage{bm}
\usepackage[breaklinks=true]{hyperref}
\usepackage[titletoc]{appendix}
\usepackage{color}
\usepackage{subfig}

\input epsf.sty

\begin{document}

\title{On the influence of a patterned substrate on crystallization in suspensions of hard spheres} 

\author{Sven Dorosz}
\affiliation{Theory of Soft Condensed Matter, Universit\'e du Luxembourg, L-1511 Luxembourg, Luxembourg}

\author{Tanja Schilling}
\affiliation{Theory of Soft Condensed Matter, Universit\'e du Luxembourg, L-1511 Luxembourg, Luxembourg}

\begin{abstract}
We present a computer simulation study on crystal nucleation and growth 
in supersaturated suspensions of mono-disperse hard spheres induced by a 
triangular lattice substrate. The main result is that compressed substrates 
are wet by the crystalline phase (the crystalline phase directly appears 
without any induction time), while for stretched substrates we observe 
heterogeneous nucleation. The 
shapes of the nucleated crystallites fluctuate strongly. 
In the case of homogeneous nucleation amorphous precursors have been 
observed (Phys.~Rev.~Lett.~{\bf 105}(2):025701 (2010)). For 
heterogeneous nucleation we do not find such precursors. The fluid is directly transformed into 
highly ordered crystallites.
\end{abstract}

\pacs{82.60.Nh, 64.60.Q-, 64.60.qe, 64.70.pv, 68.55.A-}
\maketitle
When a supersaturated fluid crystallizes, crystallization is usually 
induced by the container walls, rather than to proceed from a fluctuation 
in the bulk of the system. This effect, called heterogeneous 
nucleation, is of fundamental importance for the kinetics of phase transitions 
(such as the formation of ice in the supersaturated vapor of clouds), 
as well as for technological applications, in which the 
properties of the walls can be designed to influence the properties of the 
crystals that are formed. In this article we discuss heterogeneous crystal 
nucleation and growth from the overcompressed fluid of hard spheres.  

Hard spheres have served successfully as a simple model system for fluids 
and crystals
over the past fifty years. The interaction energy between two hard 
spheres is either infinite (if they overlap) or zero (if they do not overlap), 
thus the phase behavior of the model is purely determined by entropy. 
The simplicity of the potential makes hard spheres particularly suited for 
computer simulations; and the entropic nature of the phase transition makes 
them a useful limit case for comparison to other systems, which are 
governed by an interplay between entropy and enthalpy. 

Hard spheres are not only of interest to the theoretician, they are 
also often synthesized on the colloidal scale and used in experiments
on fundamental questions of statistical mechanics 
(see e.~g.~\cite{Poon2011} and references therein). 

As the topic of our work, crystallization of hard spheres on a substrate, 
has been studied 
experimentally \cite{Hoogenboom2003B,Hoogenboom2003C,Hoogenboom2004,
Blaaderen1997,Hermes2011, Sando2011}
and theoretically \cite{Auer2003, Dijkstra2004, Volkov2002, Xu2010, 
Cacciuto2005, Heni2001, Heni2000, Esztermann2005,Wang2007} before, 
we briefly lay out in the following, which aspects of this 
topic have been focused on in the articles cited above.

The supersaturated fluid of hard spheres in contact with a planar hard wall 
has been addressed in computer simulation studies by 
Dijkstra \cite{Dijkstra2004}, Auer \cite{Auer2003} and 
Volkov \cite{Volkov2002}. 
These studies show that the planar hard wall is wet by the crystalline 
phase, hence crystallization proceeds layer by layer rather than by 
the nucleation of crystallites. 
(For a review on wetting and film growth of crystalline phases on structured and 
unstructured surfaces in various systems, including hard spheres, see the 
article by Esztermann and L\"owen \cite{Esztermann2005}.) 
Also the recent experimental and simulation work by Sandomirski and 
co-workers \cite{Sando2011} dealt with 
the growth of a crystalline film in contact with a wall. Here the wall was 
not planar but a fcc layer of spheres. The authors
found that the speed of the crystallization front depends non-monotonically 
on the packing fraction of the fluid and that a depletion zone is present 
in front of the growing crystal.

Heterogeneous nucleation of hard sphere crystals has mainly been 
addressed in the context of template-induced crystallization. Van 
Blaaderen and co-workers
\cite{Hoogenboom2003C,Hoogenboom2004,Blaaderen1997} 
showed how to design structured templates to induce the epitaxial
growth of large monocrystals and of metastable phases in a sedimenting 
liquid of hard spheres. 
Cacciuto and Frenkel studied the effect of finite 
templates of various sizes and lattice structures on crystallite 
formation by means of computer simulation \cite{Cacciuto2005}. 
Recently this topic was taken up again and 
investigated in more detail experimentally and theoretically by the groups 
of Dijkstra and van Blaaderen \cite{Hermes2011}. For small two-dimensional 
seeds of triangular as well as square symmetry they find that nucleation 
barriers depend on the seed's symmetry as well as the lattice spacing. 
This effect is due to defects and changes in crystal morphology that are 
induced by the seed.

Heterogeneous nucleation of hard spheres on an infinite substrate 
has recently been addressed by Xu and co-workers \cite{Xu2010} in a computer 
simulation study. In this work triangular and square substrates as well as a 
hcp(1100) pattern were brought in contact with a strongly overcompressed fluid,
and the evolution of the density profile 
perpendicular to the substrate as well as the fraction of crystalline 
particles were monitored. A metastable bcc-phase that was stabilized for long times was observed.

Here we present an extended simulation study of crystallization 
mechanisms and rates for a fluid of hard spheres brought in contact 
with a triangular substrate for varying overcompression and lattice distortion. 
To our knowledge there is no systematic study on the effect that distortion 
of an infinite substrate lattice has on the crystallization mechanism and 
rate of hard spheres.

We would like to close this brief overview by pointing out that  
there are other useful model systems for crystal nucleation, as 
e.~g.~complex plasmas. In contrast to colloidal systems microscopic dynamics 
in complex plasmas are almost undamped, \cite{Zuzic2006} hence they offer 
a complementary experimental approach to the topic. 

\section{Setup of the system and simulation details}
The simulations were carried out by means of an event driven molecular 
dynamics program for fixed particle number, volume and energy
(for details on event driven MD see 
refs.~\cite{Alder1959,Krantz1996,Marin1995,Lubachevsky1991}).
We simulated $N=216,000$ hard spheres of diameter $\sigma$ in contact with a 
substrate of triangular symmetry formed by $N=4200$ spheres of the same 
diameter $\sigma$. The 
substrate particles were immobile (i.e.~they had infinite mass). 
The simulation box had periodic boundaries in x and y directions. The 
substrate layers were fixed at $z=\pm \frac{L_z}{2}$ for 
$L_z = 30\sigma \ldots 50\sigma$, depending 
on the overcompression. The initial velocities were drawn from a Gaussian 
distribution and the initial mean kinetic energy per particle was set 
to $3\;k_BT$.

To monitor crystallinity, we used the local q6q6-bond-order 
parameter \cite{Steinhardt1983,Wolde1995}, which is defined as follows:
For each particle $i$ with $n(i)$ neighbors, the local bond-orientational 
structure is characterized by
\[
\bar{q}_{6m}(i) := \frac{1}{n(i)}\sum_{j=1}^{n(i)} Y_{6m}\left(\vec{r}_{ij}\right)\quad ,
\]
where $ Y_{6m}\left(\vec{r}_{ij}\right)$ are the spherical harmonics with $l=6$. $\vec{r}_{ij}$ is the displacement between particle $i$ and its 
neighbor $j$ in a given coordinate frame.   
A vector $\vec{q}_{6}(i)$ is assigned to each particle, the 
elements $m=-6 \dots 6$  of which are defined as 
\begin{equation}
q_{6m}(i) := \frac{\bar{q}_{6m}(i)}{\left(\sum_{m=-6}^6|\bar{q}_{6m}(i)|\right)^{1/2}} \quad . \label{Defq6q6}
\end{equation}
We counted particles as neighbors if their distance satisfied 
$|\vec{r}_{ij}| < 1.4\sigma$. Two neighboring 
particles $i$ and $j$ were regarded as ``bonded'' within a crystalline region 
if $\vec{q_6}(i)\cdot \vec{q_6}(j) > 0.7$. We define $n_b(i)$ as the number 
of ``bonded'' neighbors of the $i$th particle. (In the online version we 
use the following colour-coding for the snapshots: if a particle has 
$n_b > 10$, i.e. an almost perfectly hexagonally 
ordered surrounding, it is color-coded green, if 
$n_b>5$ it is color-coded brown.)

We studied various densities between particle number density 
$\rho:= N\sigma^3/V = 1.005$ (which corresponds to a volume 
fraction $\eta = 0.5262$) and  $\rho= N\sigma^3/V = 1.02$ ($\eta = 0.5341$). 
At these densities the chemical potential difference per particle 
between the metastable fluid and the stable crystalline state is 
between $\Delta \mu \simeq -0.5\;k_BT$ and $\Delta \mu \simeq -0.54\;k_BT$. 
The overcompressed fluid configurations did not show pre-existing 
crystallites that might have been created during the preparation process.

\begin{figure}[ht!]
\begin{center}
\epsfxsize=4.35in\ \epsfbox{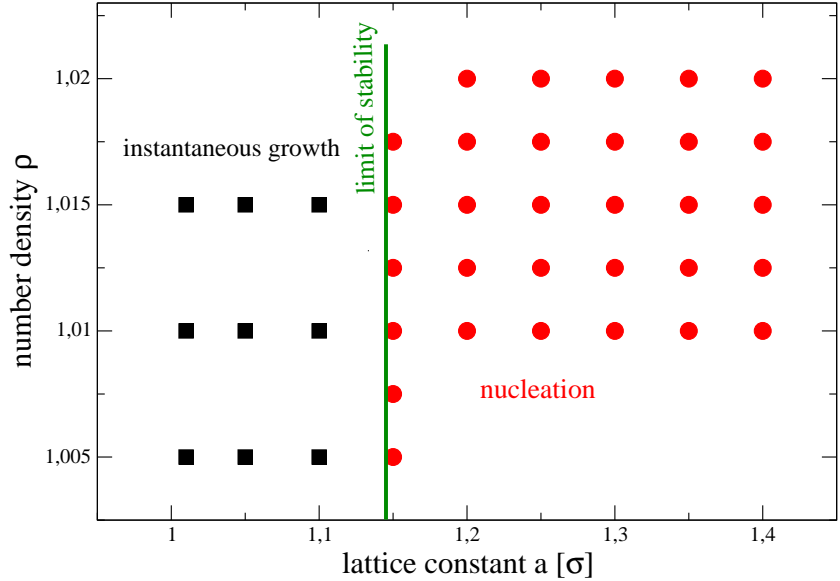}
\caption{\label{Fig1}Representation of all combinations of 
density $\rho$ and substrate lattice constant $a$ studied in this work. 
The limit of stability of the homogeneous bulk crystal 
is indicated by the solid line (green online). At substrate lattice constants 
smaller than this value (squares) we find complete 
wetting of the substrate and instantaneous film growth. Systems with a 
larger substrate lattice constant (circles) exhibit incomplete wetting and 
heterogeneous nucleation up to $a\leq 1.5\sigma$. Above this stretching, 
no heterogeneous nucleation event was observed on the scale of the simulation time.}
\end{center}
\end{figure}

Figure \ref{Fig1} shows the densities $\rho$ and substrate lattice 
constants $a$ (of the fcc-(111) plane) for which we carried out simulations.
The lattice constant indicated by the solid line (green online) corresponds 
to the bulk crystal at the spinodal, i.e.\ at the density at which 
the crystal ceases to be metastable with respect to the liquid. We obtained 
this density by simulation as well as from density functional 
theory \cite{Oettel}. The corresponding lattice constant is $a_{\rm sp}=1.15\sigma$ 
(DFT) resp. $a_{\rm sp}=1.14\sigma $ (simulation). 
One result of our study is that this line 
separates the parameter space into regions of different crystallization 
mechanisms. For $a<a_{\rm sp}$, we observed the instantaneous formation of a 
film, which then grew with time. For $a>a_{\rm sp}$, the system crystallized via heterogeneous nucleation. The transition between the two mechanisms seems to be continuous. For $a\geq 1.5\sigma$ no heterogeneous nucleation event was observed on the scale of the simulation time.

\section{Complete wetting of the substrate}
\begin{figure}[ht!]
\begin{center}
\begin{tabular}{cc}
\epsfxsize=8cm \epsfbox{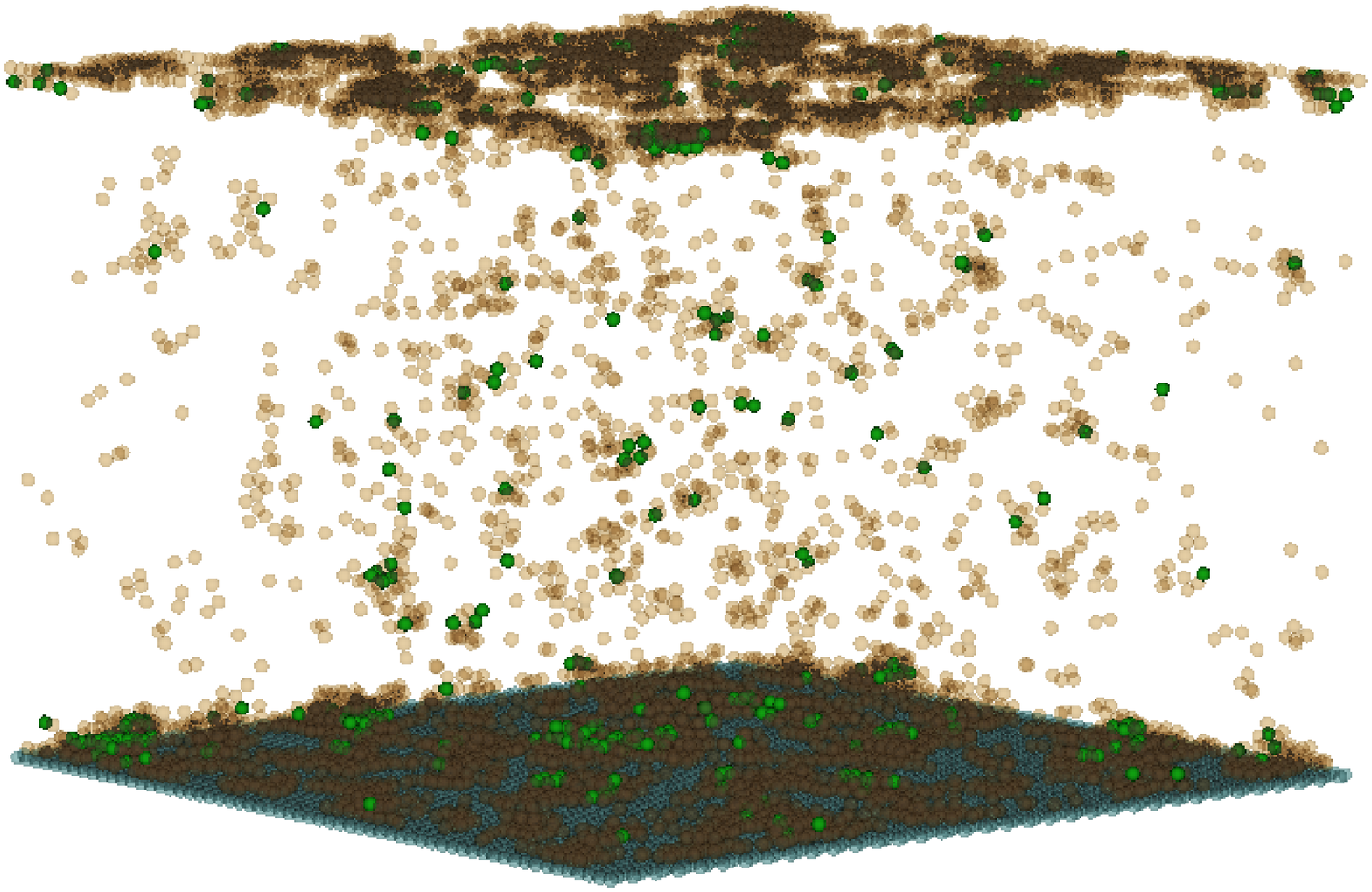}&\epsfxsize=8cm \epsfbox{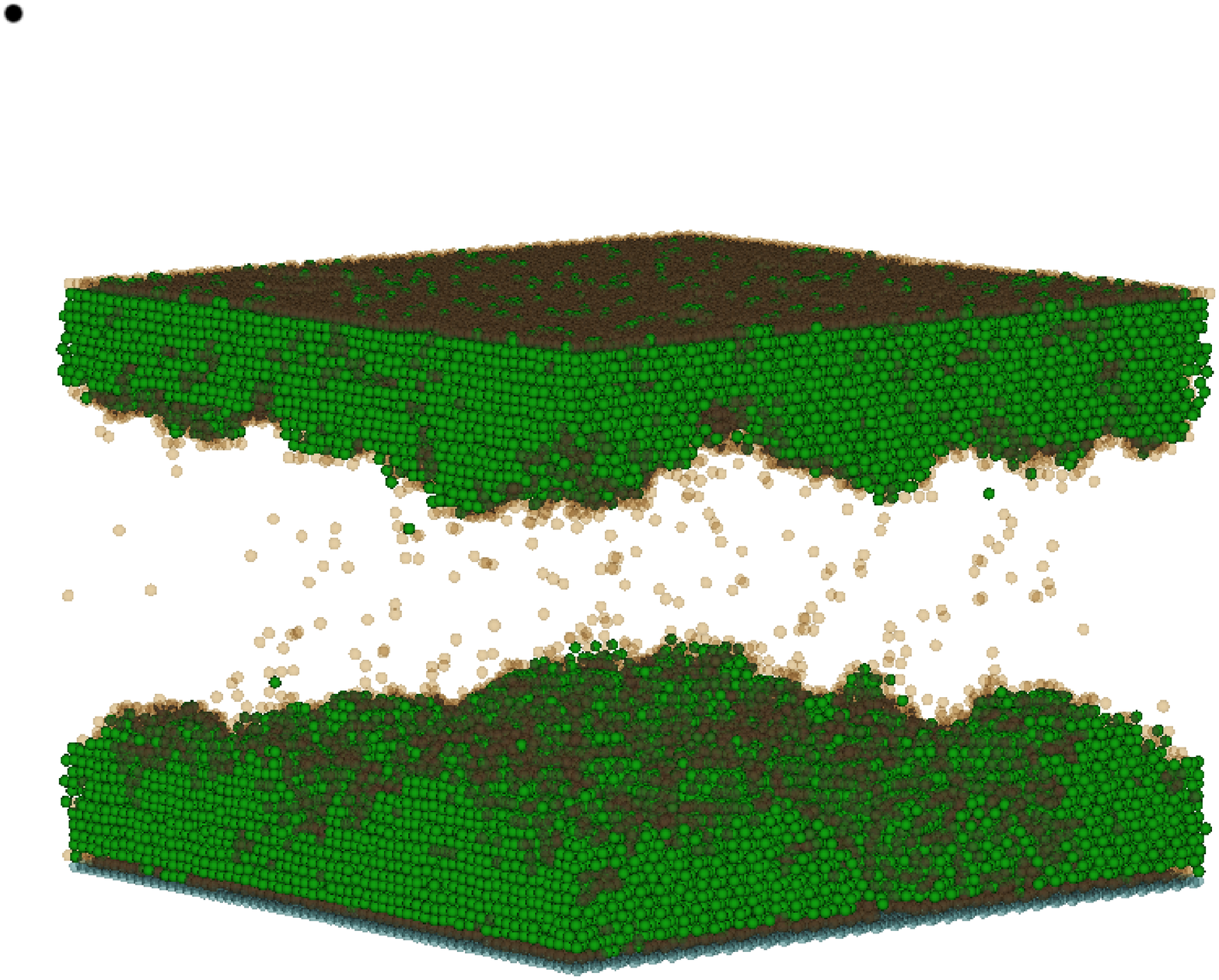}
\end{tabular}
\end{center}
\caption{\label{Fig2}Snapshots $t=\tau$ (left) and $t=100\tau$ after bringing the 
overcompressed fluid in contact with the substrate, $a=1.1\sigma$ (slightly less
than $a_{\rm sp}$), $\rho=1.01$. Only crystalline particles are shown ($n_b > 5$)}
\end{figure}

For all compressed substrates ($a < a_{\rm sp}$) we observed the formation and 
growth of a crystalline film. Typical snapshots are presented in 
figure \ref{Fig2}. (Here, we chose a system at $a=1.1\sigma$, close to 
$a_{\rm sp}$, and a bulk density of $\rho=1.01$.) The timescale of the MD simulation is expressed in multiples of $\tau=\frac{\sigma^2}{ 6D}$, with $D$ being the long-time self diffusion coefficient in the bulk fluid obtained in the same MD simulations. In the regime of densities analyzed, the diffusion constant varies by only $5\%$.)

In order to analyze the crystalline layers quantitatively, we computed the 
2-dimensional bond-order parameter $\psi_6$ for planes perpendicular to 
the $z$-direction. ($\psi_6$ is the 2d equivalent of $\bar{q}_6$.)

\[
\psi_6(i) := \frac{1}{n(i)}\sum_{j=1}^{n(i)} e^{i6\theta_{ij}}\quad ,
\]
where $\theta_{ij}$ is the angle of the vector $\vec{r}_{ij}$ and an 
arbitrary but fixed axis in the plane. We impose a cut-off at 
$|\vec{r}_{ij}| < 1.4\sigma$ and demand for a crystalline particle that 
$\psi_6(i)\psi^*_6(j)>0.7$ for six neighbors.\\
To discuss the analysis in detail, we pick three substrate lattice 
constants $a=\{1.01\sigma, 1.05\sigma, 1.1\sigma\}$ at a fixed density 
$\rho=1.005$.

\begin{figure}[ht!]
\begin{center}
	\epsfxsize=10cm \epsfbox{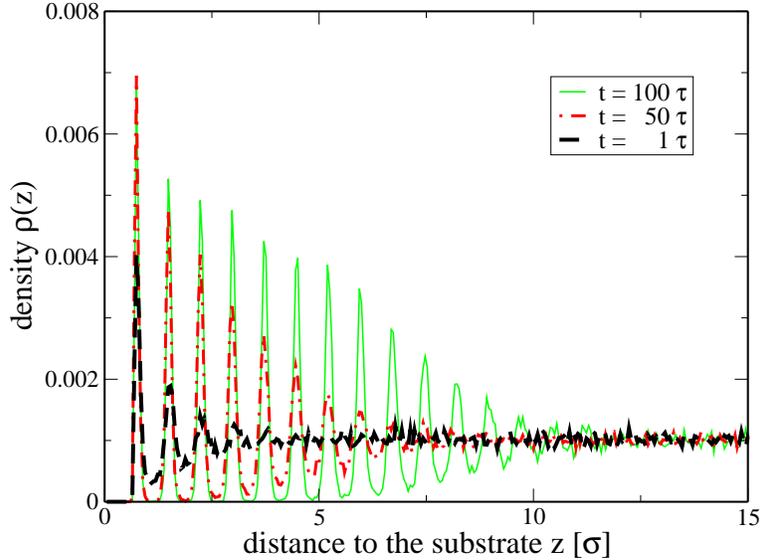}
\end{center}
\caption{\label{Fig3} Density profile perpendicular to the substrate for 
half of the system at different times. $\rho=1.01$, $a=1.1\sigma$. A film 
of layers grows.}
\end{figure}

\begin{figure}[ht!]
\begin{center}
	\epsfxsize=10cm \epsfbox{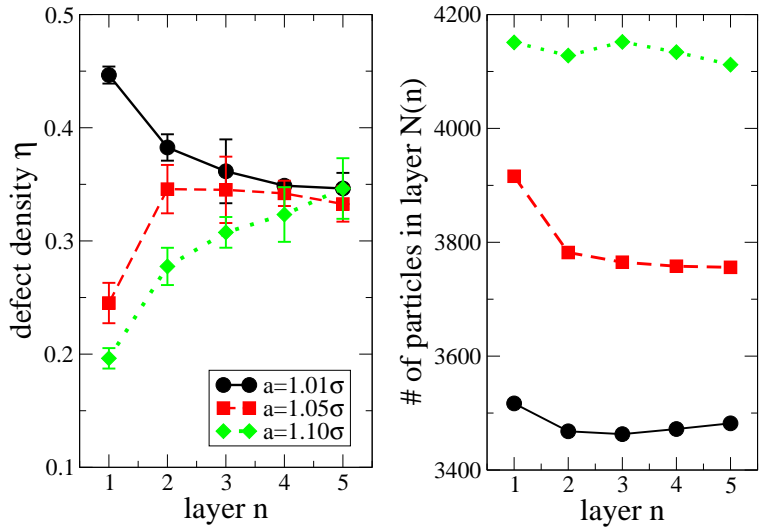}
\end{center}
\caption{\label{Fig4}(left) Defect density as a function of the index of 
each layer counted from the substrate for three different substrate 
lattice constants. The data shown has been obtained in the long time limit 
$t>400\;\tau$ and it is averaged over three independent runs each. (right) 
Number of particles $N(n)$ in each layer $n$.}
\end{figure}

Figure \ref{Fig3} shows a vertical density profile.
As a function of time the layering becomes more pronounced, as seen 
from the growth of the maxima and the appearance of voids in between 
the layers. (A quantitative analysis of the growth rate for different 
substrate lattice constants is not reported, because the lateral dimension was too small.)  
According to these profiles we identify the particles that belong to a given 
layer and study the hexagonal structure in the plane.
The overall defect density $\eta$ in a given layer n with a total numer of $N(n)$ particles is defined as 
\begin{equation}
 \eta(n):=\frac{N(n)-N_{{\rm crys}(n)}}{N(n)},
\end{equation}
with $N_{\rm crys}(n)$ being the number of crystalline particles in layer $n$.
The analysis of the defect density is shown in figure \ref{Fig4}. We have also 
included the total number of particles $N(n)$ in each layer n for the three 
cases of $a$. The further the substrate is compressed with respect to the 
equilibrium lattice the larger is the defect density in the first layer. With larger distance from the 
substrate the defect density for all three values of $a$ converges to a substrate 
independent value. At this point stresses induced by the substrate 
do not play a role in the growing crystal anymore. Only the tension 
induced by the 
shape of the periodic box, which is not commensurate with the equilibrium 
lattice, matters.

\begin{figure}[ht!]
\begin{center}
\begin{tabular}{cc}
\epsfxsize=8cm \epsfbox{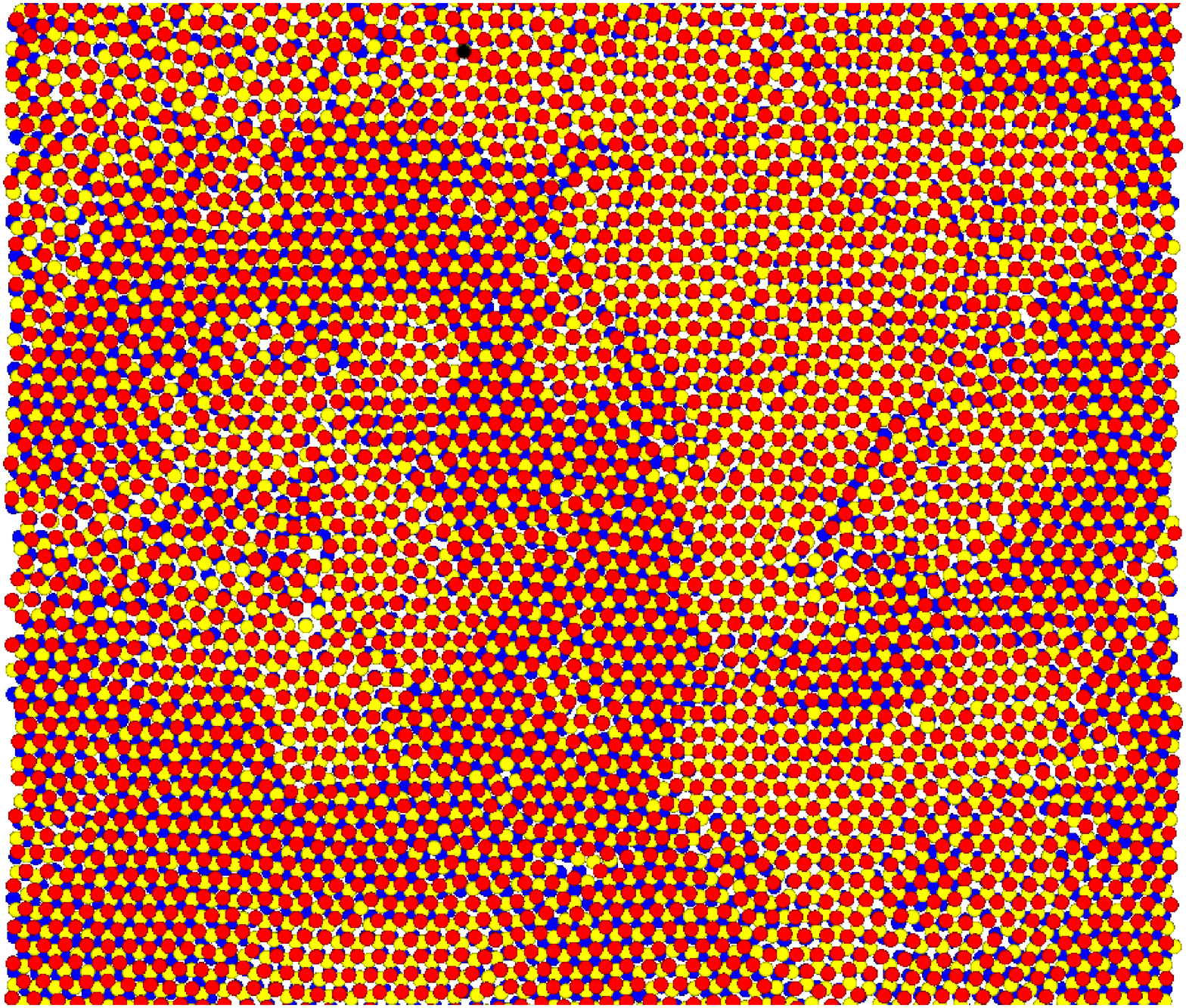}&\epsfxsize=8cm \epsfbox{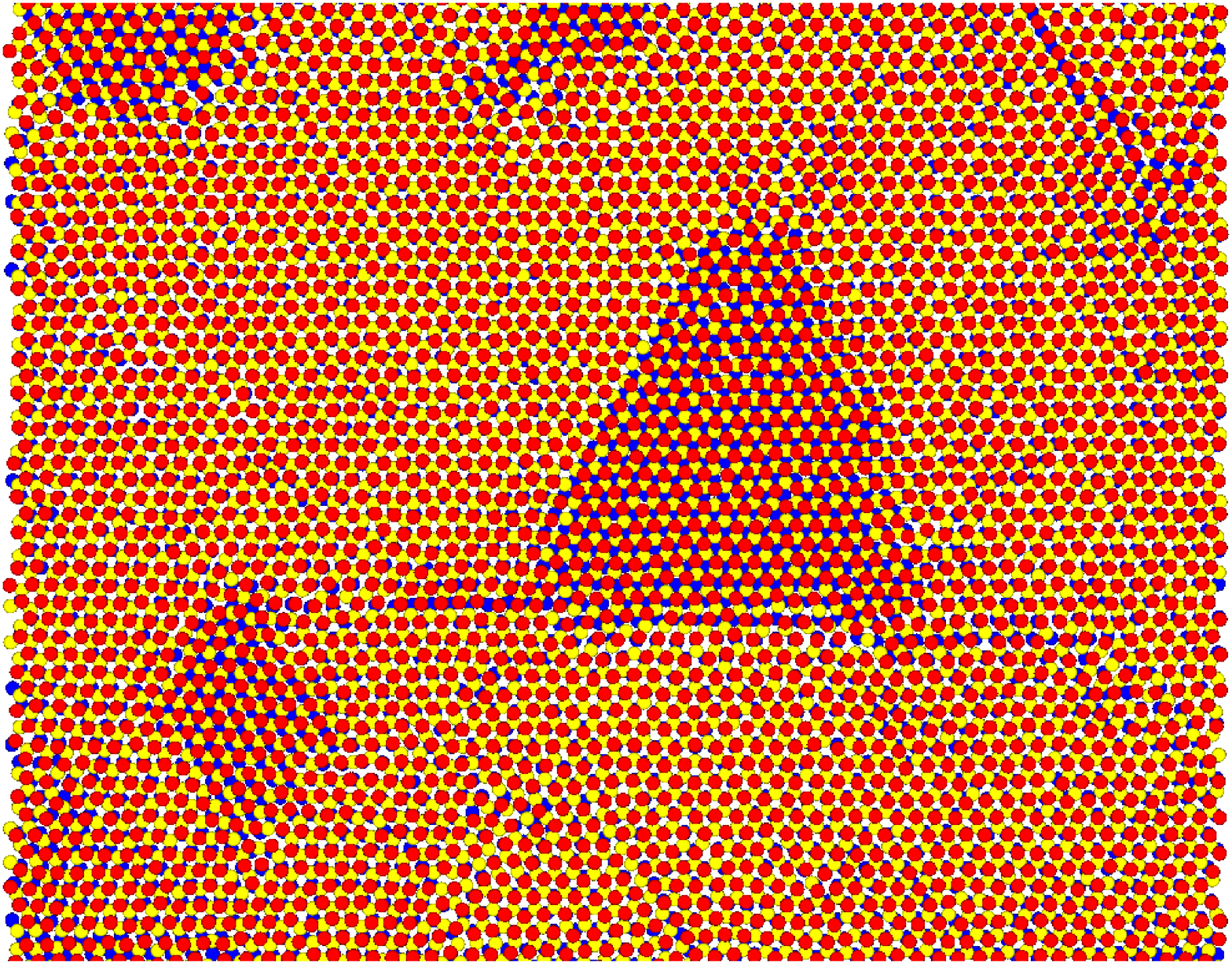}
\end{tabular}
\caption{\label{Fig5} Snapshot of the first three layers on top of the substrate for (left) $a=1.01\sigma$ and (right) $a=1.10\sigma$. The snapshots correspond to the data analyzed in figure \ref{Fig4}. There is no preference of fcc over hcp.}
\end{center}
\end{figure}

Figure \ref{Fig5} shows the covering of the substrate for the first three 
layers after $t=400\;\tau$. There is no preference of 
fcc over hcp. An analysis of the subsequent layers showed that the stacking 
is random-hcp. This is in agreement with the small free energy difference of $26\pm 6\cdot 10^{-5} k_B T/\sigma^2$ per particle \cite{Pronk1999}. Domains of equal structure are much larger for the case 
$a=1.1\sigma$ than for $a=1.01\sigma$, where there are more domain walls. No 
single crystal phase evolved on the recorded timescales. 

\section{Heterogenous nucleation near the substrate}
\begin{figure}
\begin{center}
\subfloat[$t=50\;\tau$]{\includegraphics[width=0.45\textwidth]{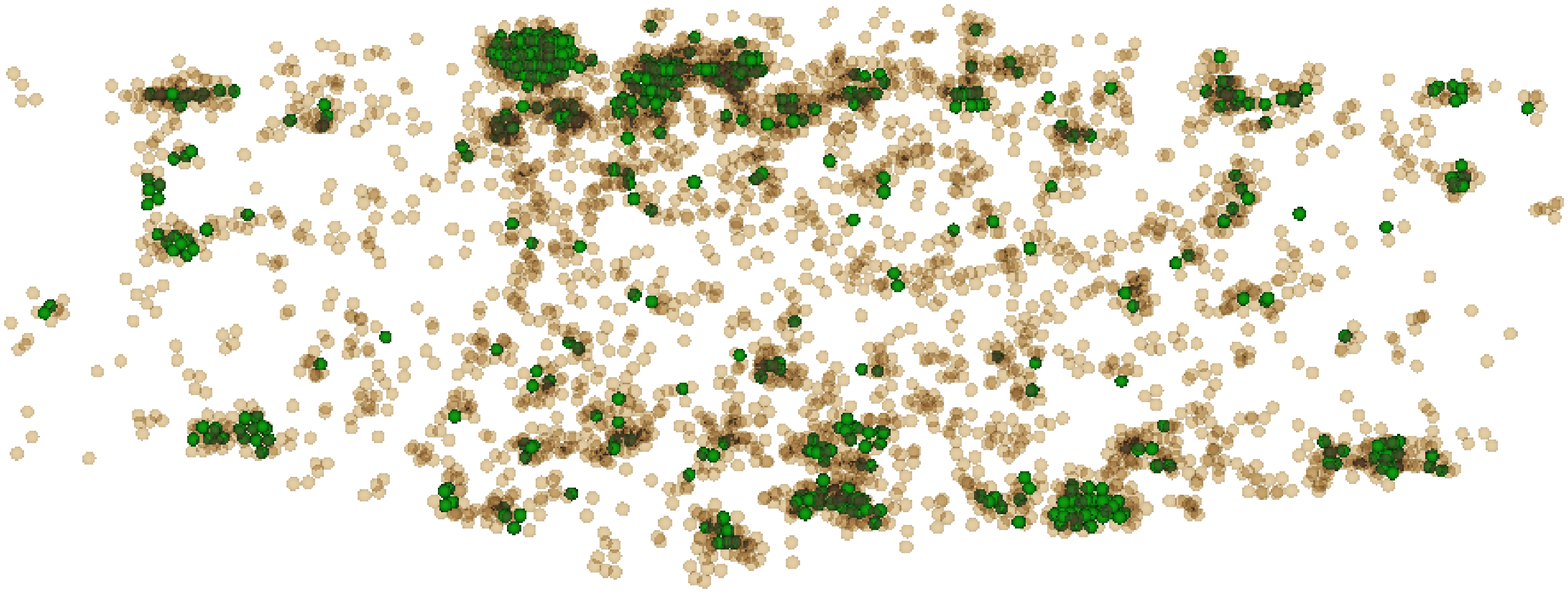}}
\subfloat[$t=150\;\tau$]{\includegraphics[width=0.45\textwidth]{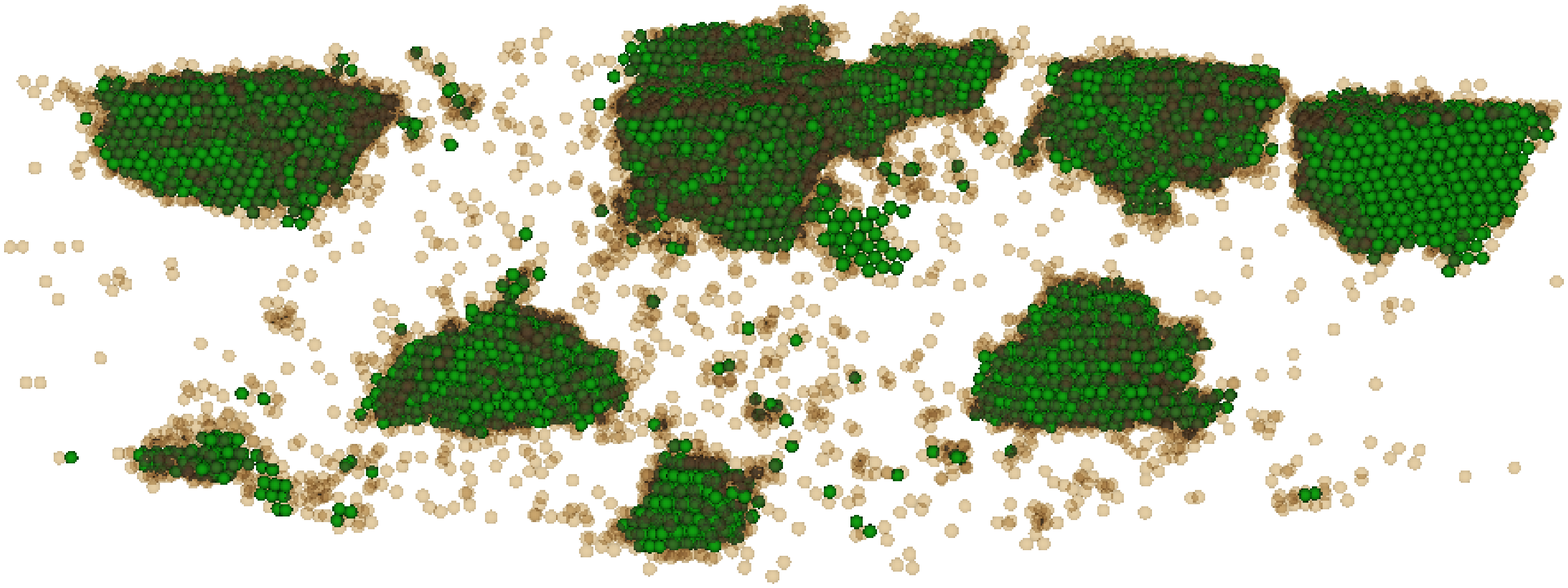}}

\caption{\label{Fig6}Snapshots at different times after bringing the 
overcompressed fluid in contact with the substrate, $a=1.4\sigma$, $\rho=1.01$. Crystallite formation at the wall dominates the nucleation process. For clarity, we are not showing the substrate. Figure (a) shows the nucleation event at which the first crystallite reaches 100 solid particles. Figure (b) shows the state of the system at a much later time.}
\end{center}
\end{figure}

For the parameter regime $1.15\sigma \leq a \leq 1.4\sigma$ we 
observe the formation of 
crystallites at the substrate. Figure \ref{Fig6} shows snapshots of 
typical crystallites at the first nucleation event (figure \ref{Fig6}a) and at a much later time 
(figure \ref{Fig6}b).\\

We define the nucleation event as the moment when the first crystalline cluster reaches a size of 100 particles, see figure \ref{Fig6}a for a snapshot. In all simulations we observed irreversible growth above this threshold. Below this threshold crystallites appeared and decayed again. Changing this value by $\pm 10$ particles does not affect any of the results presented in the following.

\begin{figure}[ht!]
\begin{center}
	\epsfxsize=8cm \epsfbox{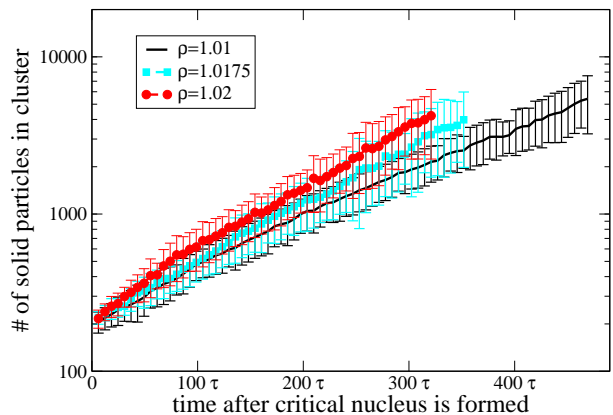}
\end{center}
\caption{\label{Fig7} Time evolution of size of the largest cluster for varying density $\rho$. The data is averaged over 8 independent runs at each given density.}
\end{figure}

In figure \ref{Fig7}, we show that the mean size of the largest crystallite can be described by a growth law that is approximately exponential with time once the
nucleation event has set in. (The timescale is reset to the nucleation event for each simulation run to compare the growth law. For each pair of $a$ and $\rho$ all data shown here is averaged over 8 independent runs.)

\begin{figure}[ht!]
\begin{center}
	\epsfxsize=8cm \epsfbox{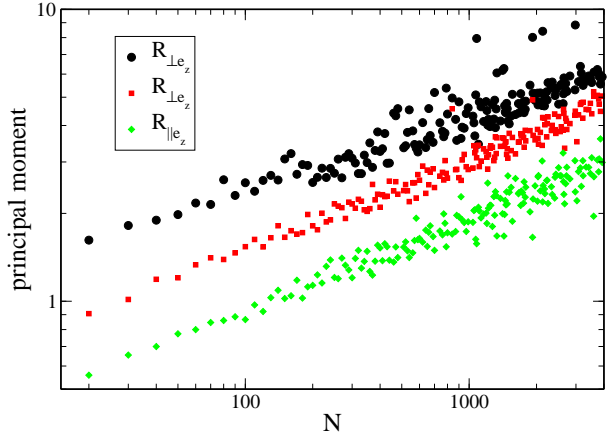}
\end{center}
\caption{\label{Fig8}Eigenvalues of the gyration tensor for all crystallites
observed during the simulation at $a=1.4\sigma$ and $\rho=1.01$. The data 
is plotted independent of time as 
a function of the number of solid particles in the crystallite. It was 
checked that the eigenvector of the smallest eigenvalue is perpendicular 
to the substrate surface.}
\end{figure}

The structure of the nuclei is analyzed by means of the tensor of gyration, 
which we diagonalized to obtain the principal moments. We identified the 
eigenvector with the smallest eigenvalue $\vec{e}_{{\rm small}} $ and checked that 
it was parallel to the substrate normal 
( $ \vec{e}_{{\rm small}}\cdot\vec{e_z}>0.9 $ 
is satisfied by more than $90\%$ of the 
crystallites, however deviations are stronger for small crystallites of 
less than 50 particles). 
As a function of the total number of particles in a crystallite 
we present the principal moments in figure \ref{Fig8}. Even up to 
$N_{\rm crys}=4000$ the 
statistics do not support the interpretation of the droplets growing as a 
spherical cap (or any other simple geometry) on the substrate. 
They are rather ramified instead.

In the case of homogeneous nucleation from the overcompressed bulk fluid in 
hard spheres, a process mediated by amorphous precursors has been 
observed \cite{Schilling2010, Schilling2011}.
We carefully checked the heterogeneous nucleation data  
and did not find any evidence of such precursors of low crystalline symmetry. 
Nucleation at the substrate immediately produces highly ordered crystallites.
Presumably the orientational symmetry breaking due to the substrate suffices to 
significantly reduce the induction time needed to create bond-orientational 
order. 

\begin{table}[htdp]
\caption{\label{table_rates}Nucleation rates for different substrate lattice spacings $a$ and densities $\rho$. All rates averaged over 8 runs. The rates are given in units of $6D/\sigma^5$. }
\begin{center}
\begin{tabular}{|c|c|c|c|c|c|c|}
\hline
$\rho \setminus  a$ & 1.15$\sigma$& 1.20$\sigma$& 1.25$\sigma$ &1.30$\sigma$ &1.35$\sigma$ &1.40$\sigma$\\
\hline
1.005 & $1.3\pm 0.2\cdot 10^{-5}$ &&&&& \\
1.0075 & $1.4\pm 0.2\cdot 10^{-5}$&&&&&\\
1.01 & $1.4\pm 0.2\cdot 10^{-5}$& $1.1\pm 0.2\cdot 10^{-5}$& $ 7.5\pm 0.7\cdot 10^{-6} $& $ 4.8\pm 0.5\cdot 10^{-6} $& $ 3.7\pm 0.5\cdot 10^{-6} $& $ 2.0\pm 0.4\cdot 10^{-6} $\\
1.0125 & $1.5\pm 0.2\cdot 10^{-5}$& $1.1\pm 0.1\cdot 10^{-5}$& $ 7.9\pm 0.8\cdot 10^{-6} $& $ 5.2\pm 0.6\cdot 10^{-6} $& $ 4.0\pm 0.6\cdot 10^{-6} $& $ 2.1\pm 0.5\cdot 10^{-6} $\\
1.015 & $1.5\pm 0.2\cdot 10^{-5}$& $1.2\pm 0.1\cdot 10^{-5}$& $ 8.5\pm 0.7\cdot 10^{-6} $& $ 5.6\pm 0.6\cdot 10^{-6} $& $ 4.8\pm 0.8\cdot 10^{-6} $& $ 2.3\pm 0.4\cdot 10^{-6} $\\
1.0175 & $1.6\pm 0.2\cdot 10^{-5}$& $ 1.3\pm 0.1\cdot 10^{-5} $& $ 8.8\pm 0.9\cdot 10^{-6} $& $ 6.0\pm 0.8\cdot 10^{-6} $ & $ 4\pm 1\cdot 10^{-6} $& $ 2.8\pm 0.5\cdot 10^{-6} $\\
1.02 & & 				$ 1.3\pm 0.1\cdot 10^{-5} $ & $ 9\pm 1\cdot 10^{-6} $& $ 5.9\pm 0.7\cdot 10^{-6} $& $ 4.6\pm 0.8\cdot 10^{-6} $ & $ 2.8\pm 0.4\cdot 10^{-6} $\\
\hline
\end{tabular}
\end{center}
\end{table}%

\begin{figure}[ht!]
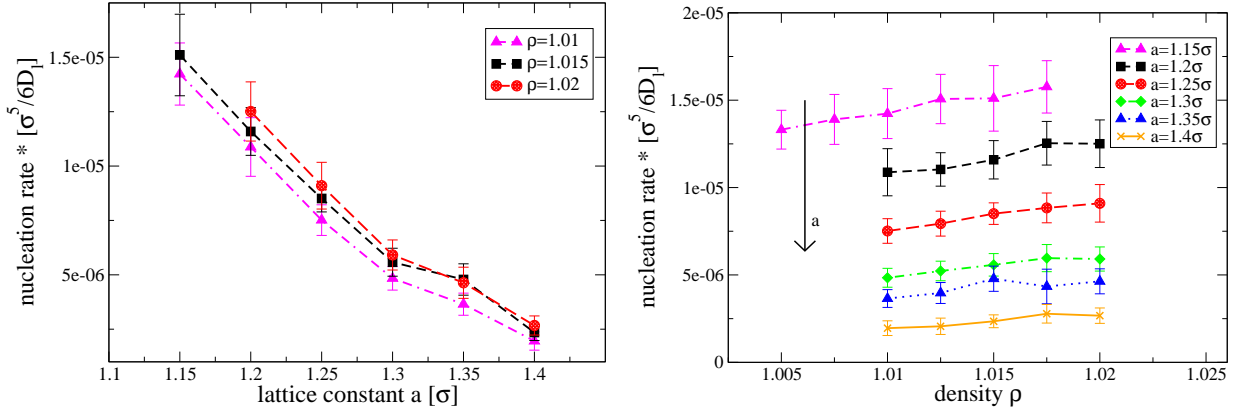

\begin{center}
\begin{tabular}{c c}
\epsfxsize=8cm\ \epsfbox{figure9a.eps}&\epsfxsize=8cm\ \epsfbox{figure9b.eps}
\end{tabular}
\caption{\label{Fig9}{(left) Nucleation rates as a function of the substrate lattice constant $a$ in the regime of droplet formation for different bulk densities $\rho$. The nucleation rates are expressed in units of $\frac{\sigma^5}{6D}$. (right) Nucleation rates as a function of the bulk density $\rho$ for different substrate lattice constants $a$.} }

\end{center}

\end{figure}

Figure \ref{Fig9}(left) shows the nucleation rates as a function of the substrate lattice constant for different bulk densities 
(also listed in table \ref{table_rates}.) We determine the nucleation rate by averaging over the times required to form the first cluster for 8 independent trajectories. (We did not include the times for subsequent events. Hence, the nucleation rates should not be affected by interactions between clusters, as they occur close to the line of stability.)  Compared to the bulk nucleation rates, (see e.~g.~ref.~\cite{Schilling2011,Filion2010} for a compilation of experimental as well as simulation results), we note that  
the heterogeneous nucleation rates are increased by several orders of 
magnitude especially at low densities. It is remarkable that the nucleation 
rates do not decrease exponentially as in the homogenous case for smaller 
densities. We rather observe, in figure \ref{Fig9}(right) a linear decrease in this 
regime of densities. 
This linear behavior is seen for all lattice constants that we analyzed. 
The slopes do not show a significant dependence on $a$.

\subsection{Conclusion}
We have studied the crystallization of an overcompressed fluid of hard spheres 
in contact with a fixed triangular lattice substrate by means of event driven 
molecular dynamics simulation. Depending on the lattice constant of the substrate, the system either 
crystallizes directly, without an induction time, or it crystallizes via nucleation. 
The value of the lattice constant that separates the two regimes is the value at which the bulk 
crystal, when being stretched, becomes unstable with respect to the liquid. If the substrate 
lattice constant is smaller than this value crystallization proceeds via 
the formation of a complete film which grows layer by layer. 
The stacking is random-hcp with a large density of defects.

If the substrate is stretched to lattice constants at which the bulk crystal is 
unstable, crystallization proceeds via heterogeneous 
nucleation. For moderate stretching, the nucleation rates are larger by 
several orders of magnitude with respect to homogeneous nucleation. The 
crystallites that are formed are very irregular in shape even when they contain 
up to several thousand particles. 

In the case of homogeneous nucleation amorphous precursors have been 
observed \cite{Schilling2010, Schilling2011}. For heterogeneous nucleation 
we do not find such precursors. The fluid is directly transformed into 
highly ordered crystallites.

\acknowledgements
We thank Hamed Maleki, Koos van Meel, Martin Oettel, and Friederike Schmid. This project has been financially supported by the DFG (SFB Tr6 and SPP1296) and by the National Research Fund, Luxembourg co-funded under the Marie Curie Actions of the European Commission (FP7-COFUND). Computer simulations presented in this paper were carried out using the HPC facility of the University of Luxembourg.


\end{document}